\documentclass[prl,twocolumn,showpacs,amsmath,amssymb,nofootinbib]{revtex4-1}

\usepackage{epsfig}
\usepackage{graphicx}
\usepackage{amsfonts}
\usepackage{amssymb}
\usepackage{bm}
\usepackage{latexsym}
\usepackage{amsmath}
\usepackage{hyperref}
\usepackage{xcolor}
\definecolor{rossoCP3}{cmyk}{0,0.88,0.77,0.40}

\usepackage{bbold}

\newcommand{\beq}{\begin{equation}}
\newcommand{\eeq}{\end{equation}}
\newcommand{\bea}{\begin{eqnarray}}
\newcommand{\eea}{\end{eqnarray}}

\begin{document}

%%%%%%%%%%%%%%%%%%%%%%%%%%%%%%%%%%%%%%%%%%%%%%%%%%%%%%%%%%%%%%%%%%%%

\title{\color{rossoCP3} Preheating after technicolor inflation}

\author{Phongpichit Channuie}
\email{channuie@gmail.com}
\affiliation{Physics Division, School of Science, Walailak University, \\Nakhon Si Thammarat 80160, Thailand}

\author{Peeravit Koad}
\email{harrykoad@gmail.com}
\affiliation{Computational Science Program, School of Science, Walailak University, Nakhon Si Thammarat 80160, Thailand}

%%%%%%%%%%%%%%%%%%%%%%%%%%%%%%%%%%%%%%%%%%%%%%%%%%%%%%%%%%%%%%%%%%%%
\begin{abstract}

We investigate the particle production due to parametric resonances in model of inflation where the lightest composite state stemming from the minimal walking technicolor theory plays the role of the inflaton. For model of inflation, the effective theory couples non-minimally to gravity. Regarding the preheating, we study in details a model of a composite inflaton field $\phi$ coupled to another scalar field $\chi$ with the interaction term $g^{2}\phi^{2}\chi^{2}$. Particularly, in Minkowski space, the stage of parametric resonances can be described by the Mathieu equation. Interestingly, we discover that broad resonances can be typically achieved and potentially efficient in our model causing the number of particle density in this process exponentially increases. 
\\[.03cm]
\end{abstract}
\date{\today}
\maketitle
\vskip 4pt

%%%%%%%%%%%%%%%%%%%%%%%%%%%%%%%%%%
\section{I. Introduction}
\label{intro}
%%%%%%%%%%%%%%%%%%%%%%%%%%%%%%%%%%
Inflation \cite{Starobinsky:1980te,Mukhanov:1981xt,Guth:1980zm,Linde:1981mu,Albrecht:1982wi} marks nowadays an inevitable ingredient when studying the very early evolution of the universe. The reason stems from the fact that it solves most of the puzzles that plague the standard Big Bang theory, and simultaneously is consistent with the observations \cite{Peiris:2003ff,Ade:2015lrj}. In other words, it not only gives sensible explanations for the horizon, flatness, and relic abundant problems, but also provides us primordial density perturbation as seeds of the formation for a large-scale structure in the universe. Most of the inflationary scenario proposed so far requires the introduction of new degree of freedom, e.g. a Higgs field \cite{Bezrukov:2007ep}, a gauge field \cite{Maleknejad:2011sq}, and so on, to successfully drive inflation. Yet there are other well-motivated inflationary models, e.g. a 3-form field inflation \cite{Germani:2009iq,Koivisto:2009sd} and multi-field inflation \cite{Kaiser:2013sna}. However, the underlying nature of theory of inflaton is still an open question. According to the cosmic frontier, the construction of inflationary models raises a lot of interest. Recent investigations revolutionize the possibility of model building in which the inflaton needs not be an elementary degree of freedom \cite{Channuie:2011rq,Bezrukov:2011mv,Channuie:2012bv}.

In the standard picture of the early universe, the universe passes through the period of reheating. At this stage, (almost) all elementary particles populating the universe were created. The instructive idea of mechanism for reheating was proposed, for instance, by the author of \cite{Linde:2005ht} in which reheating occurs due to particle production by the oscillating scalar field. In the simplest version, this field is the inflaton that exponentially drives the expansion of the universe. However, a phenomenological description of the reheating mechanism was first implemented in \cite{Albrecht:1982}. The authors added various friction terms to the equation of motion of the scalar field in order to imitate energy transfer from the inflaton to matter fields. However, it can be questioned what kind of terms should be practically added. The theory of reheating in application to the new inflationary scenario was further developed in refs. \cite{Abbott:1982hn}, in application to R2 inflation in ref. \cite{Starobinsky:1981}, and to non-commutative inflation in ref. \cite{Perrier:2012nr}. Here the treatment was based on perturbative theory where the knowledge of quantum field theory is mandatory.

Indeed, it has been recognized that in inflationary models the first stage of reheating occur in a regime of a parametric amplification of scalars field \cite{Traschen:1990sw,Kofman:1997yn,Shtanov:1994ce,Kofman:1994rk}. This preceding evolutionary phase is called \lq\lq preheating\rq\rq\, stage of which particles are explosively produced due to the parametric resonance. There are many analytical works which examined the preheating mechanism, e.g. \cite{Greene:1997fu,Kaiser:1995fb,Son:1996uv}. For instance, the authors of \cite{Tsujikawa:1999jh} have examined the properties of resonance with non-minimally coupled scalar field ${\chi}$ in preheating phase and have found that effective resonance is possible only by the existence of a non-minimal coupling $\xi R\chi^{2}$ term with a sizeable range of parameter $\xi$. Moreover, the same authors of \cite{Tsujikawa:1999iv} later considered higher-curvature inflation models $(R+\alpha^{n}R^{n})$ allowing to study a parametric preheating of a scalar field coupled non-minimally to a spacetime curvature. Another interesting paradigm was proposed by \cite{GarciaBellido:2008ab}. In this scenario,  they studied preheating mechanism of which the standard model Higgs, strongly non-minimally coupled to gravity, plays the role of the inflaton. Consequently, they discovered that the universe does reheat through which perturbative and non-perturbative effects are mixed. Moreover, the authors of \cite{DeFelice:2012wy} have investigated the production of particles due to parametric resonances in 3-form field inflation and found that this process is more efficient compared to the result of the standard-scalar-field inflationary scenario, e.g. \cite{Kofman:1997yn,Shtanov:1994ce,Kofman:1994rk}, in which the broad resonance tends to disappear more quickly.

In this work, we study details of preheating, following the work presented in \cite{Tsujikawa:1999jh}, in an inflationary scenario in which the inflaton is the lightest composite state, which are expected to be the most important ones for collider phenomenology, stemming from the Minimal Walking Technicolor (MWT) theory which includes a new strong sector based on the ${\rm SU(2)}$ gauge group with two Dirac flavors transforming according to the adjoint representation. For model of composite inflation, the authors of \cite{Channuie:2011rq} engaged the MWT effective Lagrangian with the standard slow-roll paradigm and we call it here \lq\lq technicolor inflation\rq\rq(TI). Interestingly, one of the salient features of technicolor inflation is not only to provide a possible resolution to the well-known $\eta$-problem\footnote{The problem in which one of the inflationary slow roll parameters, denoted by $\eta$, is proportional to the inflaton mass. Therefore, if the inflaton is a fundamental scalar, this parameter receives quantum corrections. As a result, such corrections would spoil the slow roll approximation.} \cite{Copeland:1994vg} of inflationary model-building but also to allow for an inflationary phase in the early universe \cite{Channuie:2011rq}. 

This paper is organized as follows. In the next section, we briefly review our model of inflation in which the inflaton emerges as a composite state coupled non-minimally to a spacetime curvature. Adding a scalar field coupled minimally to gravity, in Sec.(III), we then introduce an analytical approach to study the preheating for model of composite inflation. In Sec.(IV), we study parametric resonances of model when a composite inflaton field $\phi$ coupled to another scalar field $\chi$ with the interaction term $g^{2}\phi^{2}\chi^{2}$. Finally, we give our findings in the last section.

%%%%%%%%%%%%%%%%%%%%%%%%%%%%%%%%%%
\section{II. Technicolor inflation: a recap}
\label{model1}
%%%%%%%%%%%%%%%%%%%%%%%%%%%%%%%%%%
The underlying gauge theory for the technicolor-inspired inflation is the SU(N) gauge group with $N_{f}=2$ Dirac massless fermions. The two technifermions transform according to the adjoint representation of SU(2) technicolor (TC) gauge group, called ${\rm SU(2)}_{\rm TC}$. Here we engaged the simplest models of technicolor known as the minimal walking technicolor (MWT) theory \cite{Sannino:2004qp,Hong:2004td,Dietrich:2005wk,Dietrich:2005jn} with the standard (slow-roll) inflationary paradigm as a template for composite inflation and name it, in short, the MCI model. In order to examine the symmetry properties of the theory, we arrange them by using the Weyl basis into a column vector, and the field contents in this case are
\begin{equation}
{\cal Q}^{a}=\begin{pmatrix}
U^{a}_{L}\\D^{a}_{L}\\-i\sigma^{2}U^{*a}_{R}\\-i\sigma^{2}D^{*a}_{R}
\end{pmatrix}\,, \label{field}
\end{equation}
where $U_{L}$ and $D_{L}$ are the left-handed techniup and technidown respectively, and $U_{R}$ and $D_{R}$ are the corresponding right-handed particles and the upper index $a=1,2,3$ is the TC index indicating the three dimensional adjoint representation. Since the ${\cal Q}$ is four component, the technifermion fields are said to be in the fundamental representation of SU(4). With the standard breaking to the maximal diagonal subgroup, the SU(4) global symmetry spontaneously breaks to SO(4). Such a breaking is driven by the formation of the following condensate:
 \begin{eqnarray}
\left\langle{\cal Q}^{\alpha}_{i}{\cal Q}^{\beta}_{j}\epsilon_{\alpha\beta}{\cal E}^{ij}\right\rangle = -2\left\langle\bar{U}_{R}U_{L}+\bar{D}_{R}D_{L}\right\rangle\,,
\label{v-tc}
\end{eqnarray}
where $i,\,j\,=1,\,.\,.\,.\,,4$ denote the components of the tetraplet of ${\cal Q}$, and $\alpha,\,\beta$ indicate the ordinary spin. The $4\times4$ matrix ${\cal E}^{ij}$ is defined in terms of the 2-dimensional identical matrix, $\mathbb{1}$, as
\begin{equation}
{\cal E}=\begin{pmatrix}
\mathbb{0} & \mathbb{1} \\ \mathbb{1} & \mathbb{0}
\end{pmatrix}\,, \label{field1}
\end{equation}
with, for example, $\epsilon_{\alpha\beta}=-i\sigma^{2}_{\alpha\beta}$ and $\left\langle U^{\alpha}_{L}U^{*\beta}_{R}\epsilon_{\alpha\beta}\right\rangle=-\left\langle\bar{U}_{R}U_{L}\right\rangle$. The connection between the composite scalar fields and the underlying technifermions can be obtained from the transformation properties of SU(4). To this end, we observe that the elements of the matrix ${\cal M}$ transform like technifermion bilinears such that 
 \begin{eqnarray}
{\cal M}_{ij}\sim {\cal Q}^{\alpha}_{i}{\cal Q}^{\beta}_{j}\epsilon_{\alpha\beta}\quad{\rm with}\quad i,\,j=1,\,.\,.\,.\,,4\,.\label{mtc}
\end{eqnarray}
The composite action non-minimally coupled to gravity can be built in terms of the matrix ${\cal M}$ in the Jordan frame as \cite{Channuie:2011rq}
\begin{eqnarray}
{\cal S}_{\rm MCI, J}=\int d^{4}x\sqrt{-g}&&\Big[-\frac{M^{2}_{\rm P}}{2}R - \frac{1}{2}\xi{\rm Tr}\left[{\cal M}{\cal M}^{\dagger}\right]R\nonumber\\&&\,\,+{\cal L}_{\rm MWT}\Big]\,,
\end{eqnarray}
where ${\cal L}_{\rm MWT}$ is the Lagrangian density of the MWT sector, see \cite{Channuie:2011rq} for more details. From the above action, the non-minimally coupled term corresponds at the fundamental level to a four-fermion interaction term coupled to the Ricci scalar in the following way:
\begin{eqnarray}
\frac{1}{2}\xi{\rm Tr}\left[{\cal M}{\cal M}^{\dagger}\right]R = \frac{1}{2}\xi\frac{({\cal Q}{\cal Q})^{\dagger}{\cal Q}{\cal Q}}{\Lambda^{4}_{\rm Ex.}}R\,,
\end{eqnarray}
where $\Lambda_{\rm Ex.}$ is a new high energy scale in which this operator generates. Here the non-minimal coupling is added at the fundamental level showing that the non-minimal coupling is well motivated at the level of the fundamental description. However, an instructive analysis of the generated coupling of a composite scalar field to gravity has been initiated in the Nambu-Jona-Lasinio (NJL) model \cite{Hill:1991jc}. With this regard, the non-minimal coupling apparently seems rather natural. Using the renormalization group equation for the chiral condensate, we find
\begin{eqnarray}
\left\langle{\cal Q}{\cal Q}\right\rangle_{\Lambda_{\rm Ex.}}\sim \left(\frac{\Lambda_{\rm Ex.}}{\Lambda_{\rm TI}}\right)^{\gamma}\left\langle{\cal Q}{\cal Q}\right\rangle_{\Lambda_{\rm TI}}\,,
\end{eqnarray}
where the subscripts indicate the energy scale at which the corresponding operators are evaluated, and basically $\Lambda_{\rm Ex.}\gg\Lambda_{\rm TI}$. If we assume the fixed value of $\gamma$ is around two the explicit dependence on the higher energy $\Lambda_{\rm Ex.}$ disappears. This is since we have ${\cal M}\sim \left\langle{\cal Q}{\cal Q}\right\rangle_{\Lambda_{\rm TI}}/\Lambda^{2}_{\rm TI}$. According to this model at the effective description, the relevant effective theory consisting of a composite inflaton ($\phi$) and its pseudo scalar partner ($\Theta$), as well as nine pseudo scalar Goldstone bosons ($\Pi^{\mathbb{A}}$) and their scalar partners ($\tilde{\Pi}^{\mathbb{A}}$) can be assembled in the matrix form such that
\begin{eqnarray}
{\cal M}=\left[\frac{\phi+i\Theta}{2}+\sqrt{2}\left(i\Pi^{\mathbb{A}}+\tilde{\Pi}^{\mathbb{A}}\right)X^{\mathbb{A}}\right]{\cal E}\,,
\end{eqnarray}
\begin{figure}[t]
\begin{center}		
\includegraphics[width=1\linewidth]{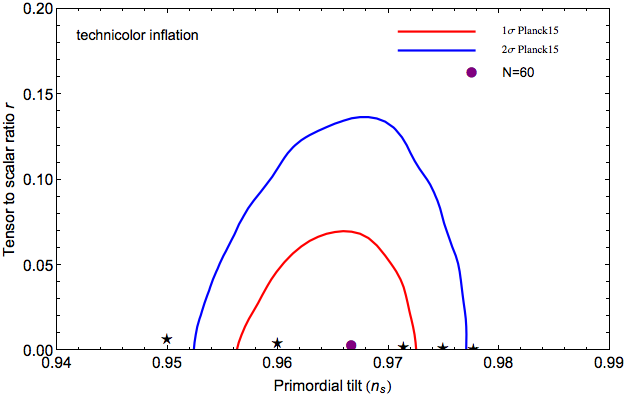}
\caption{The theoretical predictions in the $(r - n_{s})$ plane for technicolor inflation with Planck'15 results for TT, TE, EE, +lowP and assuming $\Lambda {\rm CDM} + r$ \cite{Ade:2015lrj} for different values of e-folds $40 \leq N \leq 90$. \label{tc}}
\end{center}
\end{figure}
where $X^{\mathbb{A}}$'s,\,${\mathbb{A}}=1,...,9$, are the generators of the SU(4) gauge group which do not leave the vacuum expectation value (vev) of ${\cal M}$ invariant, i.e. $\left\langle{\cal M}\right\rangle=v{\cal E}/2,\,v\equiv \left\langle\phi\right\rangle$. Here the (composite) scale of theory is identified by $\Lambda_{\rm TI}=4\pi v$, with $v$ the scale of (new) fermion condensate, implying that $\Lambda_{\rm Ex.}\gtrsim 4\pi v$. In this model, the composite inflaton is the lightest state $\phi$, and the remaining composite fields are massive. This provides a sensible possibility to consider the $\phi$ dynamics only. The relevant composite inflaton, non-minimally coupled to gravity, effective action is given by
\begin{eqnarray}
\mathcal{S}_{\rm J}=\int d^{4}x \sqrt{-g}&&\Big[- \frac{M^{2}_{\rm P}}{2}\Omega^{2}R + \frac{1}{2}(\nabla\phi)^{2}  - V(\phi) \Big]\,, \label{action00}
\end{eqnarray}
where
\begin{eqnarray}
\Omega^{2} = \frac{\left(M^{2}_{\rm P} + \xi\phi^{2}\right)}{M^{2}_{\rm P}},\,V(\phi) = -\frac{1}{2}m^{2}_{\phi}\phi^{2} + \frac{\lambda}{4}\phi^{4}\,. \label{Vphi}
\end{eqnarray}
where $\lambda$ is a self coupling and the inflaton mass is $m^{2}_{\rm TI}=2m^{2}_{\phi}$. In order to proceed our investigation below, we assume for $\phi\gg m_{\phi}/\sqrt{\lambda}$ that the inflaton mass term, $m^{2}_{\phi}\phi^{2}/2$, does not affect the frequency of oscillations of the inflaton field $\phi$. Recently, the authors of \cite{Channuie:2013lla,Channuie:2015ewa} showed that its predictions, i.e. the spectral index of curvature perturbation $n_{s}$ and the tensor-to-scalar ratio $r$, fit very well with the data from the Planck satellite \cite{Ade:2015lrj}, see Fig.\,(\ref{tc}), and from the improved analysis by the BICEP2 $\&$ Keck Array CMB polarization experiments \cite{Array:2015xqh}. 
%%%%%%%%%%%%%%%%%%%%%%%%%%%%%%%%%%
\section{III. the model}
\label{model}
%%%%%%%%%%%%%%%%%%%%%%%%%%%%%%%%%%
In this work, we will examine the preheating process for a composite model of inflation involving non-minimal coupling between gravity and matter. Many attempts have been studying inflationary models containing this kind of interaction term such as Higgs inflation \cite{Bezrukov:2007ep} and pure $\lambda\phi^{4}$ theory \cite{Futamase:1987ua,Mahajan:2013kra,Steinwachs:2013tr}. In this section, we will follow the work proposed by \cite{Kaiser:2010ps} and focus on the two-field scenario. In addition, for a simplified two-field model, there has itself been studied fairly often in the literature, e.g. see \cite{Bassett:1997az,Tsujikawa:2002nf,Amin:2014eta}. Moreover, the preheating effect of inflation for multifield scenario has been so far investigated by \cite{DeCross:2015uza}. In the following, we choose for current investigation the 4D action of our system in the Jordan (J) frame:
\begin{eqnarray}
\mathcal{S}_{\rm J}=\int d^{4}x \sqrt{-g}&&\Big[- f(\Phi^{i})R \nonumber\\&&\,\,\,\,+ \frac{1}{2}\delta_{ij}g^{\mu\nu}\nabla_{\mu}\Phi^{i}\nabla_{\nu}\Phi^{j} -V(\Phi^{i}) \Big]\,, \label{actionset}
\end{eqnarray}
where $i=1,2$ and $\Phi^{i}=(\phi,\chi)$ with $\chi$ an additional scalar field. Let us consider a typical form of $f(\Phi^{i})$ in our case in which the non-minimal couplings take the form. 
\begin{eqnarray}
f(\Phi^{i}) = \frac{1}{2}\left(M^{2}_{0} + \xi_{\Phi^{i}}(\Phi^{i})^{2}\right)\,, \label{fP}
\end{eqnarray}
where $M_{0}$ in this work is assigned to be the Planck constant $M_{\rm P}$ and the coupling strengths $\xi_{\Phi^{i}}$ are the couplings between curvature and matter fields. 

In order to bring the gravitational portion of the action into the canonical Einstein-Hilbert form, we perform a conformal transformation by rescaling $\tilde{g}_{\mu\nu} = \Omega^{2}(x)g_{\mu\nu}$. Here we can relate the conformal factor $\Omega^{2}(x)$ to the nonminimal-coupling sector via
\begin{eqnarray}
\Omega^{2}(x) = \frac{2}{M^{2}_{\rm P}}f(\Phi^{i}(x))\,, \label{con}
\end{eqnarray}
By applying the conformal transformation given above, we can eliminate the nonminimal-coupling sector and obtain the resulting action in the Einstein frame \cite{Kaiser:2010ps}
\begin{eqnarray}
\mathcal{S}_{\rm E}=\int d^{4}x \sqrt{-g}&&\Big[- \frac{M^{2}_{\rm P}}{2}R \nonumber\\&&\,\,\,\,+ \frac{1}{2}{\cal G}_{ij}g^{\mu\nu}\nabla_{\mu}\Phi^{i}\nabla_{\nu}\Phi^{j} -{\cal U}(\Phi^{i})\Big]\,, \label{ef}
\end{eqnarray}
with ${\cal U}(\Phi^{i})\equiv V(\Phi^{i})/\Omega^{4}$. Here we have dropped the tildes for convenience and ${\cal G}_{ij}$ is given by
\begin{eqnarray}
{\cal G}_{ij} = \frac{M^{2}_{\rm P}}{2f}\delta_{ij} + \frac{3}{2}\frac{M^{2}_{\rm P}}{f^{2}}f_{,i}f_{,j}\,, \label{Gij}
\end{eqnarray}
where $f_{,i}=\partial f/\partial\Phi^{i}$. In our case, the above quantity can be explicitly recast in terms of the fields $(\phi,\chi)$ as
\begin{eqnarray}
{\cal G}_{\phi\phi} &=& \frac{M^{2}_{\rm P}}{2f}\left(1+\frac{3\xi_{\phi}\phi^{2}}{f}\right)\,, \label{pp}\\
{\cal G}_{\phi\chi} &=& {\cal G}_{\chi\phi} = \frac{M^{2}_{\rm P}}{2f}\left(\frac{3\xi_{\phi}\xi_{\chi}\phi\chi}{f}\right)\,, \label{pc}\\
{\cal G}_{\chi\chi} &=& \frac{M^{2}_{\rm P}}{2f}\left(1+\frac{3\xi_{\chi}\phi^{2}}{f}\right)\,.\label{cc}
\end{eqnarray}
The action in terms of the fields $(\phi,\chi)$ takes the form
\begin{eqnarray}
\mathcal{S}_{\rm E}=\int d^{4}x \sqrt{-g}&&\Big[- \frac{M^{2}_{\rm P}}{2}R \nonumber\\&&\,\,\,\,+ \frac{1}{2}\frac{M^{2}_{\rm P}}{2f}\left(1+\frac{3\xi_{\phi}\phi^{2}}{f}\right)g^{\mu\nu}\nabla_{\mu}\phi\nabla_{\nu}\phi \nonumber\\&&\,\,\,\,+ \frac{1}{2}\frac{M^{2}_{\rm P}}{f}\left(\frac{3\xi_{\phi}\xi_{\chi}\phi\chi}{f}\right)g^{\mu\nu}\nabla_{\mu}\phi\nabla_{\nu}\chi \nonumber\\&&\,\,\,\,+ \frac{1}{2}\frac{M^{2}_{\rm P}}{2f}\left(1+\frac{3\xi_{\chi}\chi^{2}}{f}\right)g^{\mu\nu}\nabla_{\mu}\chi\nabla_{\nu}\chi \nonumber\\&&\,\,\,\,-{\cal U}(\phi,\chi)\Big]\,. \label{actionE2}
\end{eqnarray}
Notice that the model being currently investigated coincides with that of the two-field scenario in which each of them non-minimally coupled to the spacetime curvature. In such a case, there is no conformal transformation that can bring both the gravitational sector and the kinetics terms of the scalar sector into the canonical form. As a result, the fields would not behave as they would in a minimally coupled scenario.

However, in the present work we suppose that the field $\phi$ is non-minimally coupled to gravity and we set $\xi_{\phi}=\xi$; whilst only the field $\chi$ is minimally coupled to gravity, i.e. $\xi_{\chi}=0$. Therefore, the action in terms of the fields $(\phi,\chi)$ becomes
\begin{eqnarray}
\mathcal{S}_{\rm E}=\int d^{4}x \sqrt{-g}&&\Big[- \frac{M^{2}_{\rm P}}{2}R \nonumber\\&&\,\,\,\,+ \frac{1}{2}\frac{M^{2}_{\rm P}}{2F}\left(1+\frac{3\xi\phi^{2}}{F}\right)g^{\mu\nu}\nabla_{\mu}\phi\nabla_{\nu}\phi \nonumber\\&&\,\,\,\,+ \frac{1}{2}\frac{M^{2}_{\rm P}}{2F}g^{\mu\nu}\nabla_{\mu}\chi\nabla_{\nu}\chi \nonumber\\&&\,\,\,\,-{\cal U}(\phi,\chi)\Big]\,, \label{actionE3}
\end{eqnarray}
where $F \equiv f_{\xi_{\chi}=0}$. Notice that the resulting action can then be translated to a canonical form by imposing the rescaled fields. This implementation can be achieved by introducing the rescaled fields $\hat{\phi}(\phi,\chi)$ and $\hat{\chi}(\phi,\chi)$ such that
\begin{align}
\frac{\partial\hat{\phi}}{\partial\phi} = \sqrt{\frac{M^{2}_{\rm P}}{2F}\left(1+\frac{3\xi\phi^{2}}{F}\right)}\quad{\rm and}\quad \frac{\partial\hat{\chi}}{\partial\chi} = \sqrt{\frac{M^{2}_{\rm P}}{2F}}\,. \label{resca12}
\end{align}
In terms of the new fields, the action takes the form
\begin{eqnarray}
\mathcal{S}_{\rm E}=\int d^{4}x \sqrt{-g}&&\Big[ - \frac{M^{2}_{\rm P}}{2}R +\frac{1}{2}(\nabla\hat{\phi})^{2} + \frac{1}{2}(\nabla\hat{\chi})^{2} \nonumber\\&&\,\,\,\,- {\cal U}\left(\hat{\phi}(\phi,\chi),\hat{\chi}(\phi,\chi)\right)\Big]\,. \label{actionEi}
\end{eqnarray}
In this case, the second term of Eq.(\ref{resca12}) can then be integrated to yield
\begin{eqnarray}
\chi = \left(1 + \frac{\xi\phi^{2}}{M^{2}_{\rm P}}\right)^{1/2}\hat{\chi} \,. \label{tilchi}
\end{eqnarray}
To find the explicit form of the potential in terms of a new variable $\hat{\phi}$, we must verify the expression of $\phi$ in terms of $\hat{\phi}$. As suggested in \cite{GarciaBellido:2008ab}, this can simply implemented by integrating the first term of Eq.(\ref{resca12}), whose general solution is given by
\begin{eqnarray}
\frac{\sqrt{\xi}}{M_{\rm P}}\hat{\phi}(\phi) = &&\sqrt{1+6\xi}\sinh^{-1}\left(\sqrt{1+6\xi}\,{\mathbb u}\right) \nonumber\\&&- \sqrt{6\xi}\sinh^{-1}\left(\sqrt{6\xi}\frac{{\mathbb u}}{\sqrt{1+{\mathbb u}^{2}}}\right) \,, \label{tilphigen}
\end{eqnarray}
where ${\mathbb u}\equiv \sqrt{\xi}\phi/M_{\rm P}$. It is worth noting that in our case since $\xi\gg1$ and using the identity $\sinh^{-1}{\mathbb x}=\ln({\mathbb x}+\sqrt{{\mathbb x}+1})$ for $-\infty<{\mathbb x}<\infty$, this therefore allows us to rewrite the above general solution as
\begin{eqnarray}
\frac{\sqrt{\xi}}{M_{\rm P}}\hat{\phi}(\phi) \approx \sqrt{6\xi}\ln\left(1+\frac{\xi\phi^{2}}{M^{2}_{\rm P}}\right)^{1/2}  \,, \label{tilphi}
\end{eqnarray}
or, equivalently,
\begin{eqnarray}
\frac{\xi}{M^{2}_{\rm P}}\phi^{2}(\hat{\phi})\approx 1 - \exp\left(\sqrt{\frac{2}{3}}\frac{\hat{\phi}}{M_{\rm P}}\right) \,. \label{til2hatphi}
\end{eqnarray}
From the above relations, we also find
\begin{eqnarray}
\Omega^{2}(\hat{\phi}) = \exp\left(\sqrt{\frac{2}{3}}\frac{\hat{\phi}}{M_{\rm P}}\right)\,.\label{omegat}
\end{eqnarray}
Notice that the field $\hat{\phi}$ is directly related to the conformal transformation factor $\Omega$.  With the above machinery, the resulting action in the Einstein frame yields the following equations of motion for $\phi$ and $\chi$, respectively:
\begin{eqnarray}
&& \ddot{\hat{\phi}}+3\frac{\dot{a}}{a}\dot{\hat{\phi}}- \frac{1}{a^{2}}\nabla^{2}\hat{\phi}+\frac{\partial {\cal U}}{\partial\hat{\phi}} = 0 \,, \label{EoMphi}\\
&& \ddot{\hat{\chi}}+3\frac{\dot{a}}{a}\dot{\hat{\chi}} - \frac{1}{a^{2}}\nabla^{2}\hat{\chi}+\frac{\partial {\cal U}}{\partial\hat{\chi}} = 0 \,. \label{EoMchi}
\end{eqnarray}
Here we are interested in a preheating effect after inflation. To begin with, we shall assume that the spacetime and the inflaton $\phi$ give a classical background and the scalar field $\chi$ is treated as a quantum field on that background. In the next section, we will examine preheating mechanism of inflationary model which were proposed so far in the literature \cite{Channuie:2011rq}.
%%%%%%%%%%%%%%%%%%%%%%%%%%%%%%%%%%
\section{IV. parametric resonance}
\label{para}
%%%%%%%%%%%%%%%%%%%%%%%%%%%%%%%%%%
In terms of the component fields, after performing the conformal (Weyl) transformation, the transformed potential, $U\left(\hat{\phi},\hat{\chi}\right)$, is given by
\begin{eqnarray}
U\left(\hat{\phi},\hat{\chi}\right) &=& \exp\left(-2\sqrt{\frac{2}{3}}\frac{\hat{\phi}}{M_{\rm P}}\right)\Big[\frac{\lambda}{4}\phi^{4} -\frac{1}{2}m^{2}_{\phi}\phi^{2} \nonumber\\&&\quad\quad\quad\quad\quad\quad+\frac{1}{2}g^{2}\phi^{2}\chi^{2} - \frac{1}{2}m^{2}_{\chi}\chi^{2} \Big]\,, \label{actionEPott}
\end{eqnarray}
After relabelling $\hat{\phi}$ to $\phi$,\,$\hat{\chi}$ to $\chi$,\,etc., the resulting potential takes the form
\begin{eqnarray}
U\left(\phi,\chi\right) \simeq  &&\frac{\lambda M^{2}_{\rm P}}{4\xi^{2}}\left(1 - e^{-\sqrt{2/3}\phi/M_{\rm P}}\right)^{2} - \frac{1}{2}m^{2}_{\chi}e^{-2\sqrt{2/3}\phi/M_{\rm P}}\chi^{2} \nonumber\\&&+\frac{g^{2}M^{2}_{\rm P}}{2\xi}e^{-2\sqrt{2/3}\phi/M_{\rm P}}\left(e^{\sqrt{2/3}\phi/M_{\rm P}} - 1\right)\chi^{2}\,, \label{actionEPot}
\end{eqnarray}
where, in accord with our examination, we have neglected the inflaton mass term. The Klein-Gordon equation for the inflaton field is approximately given by
\begin{eqnarray}
\ddot{\phi}+3H\dot{\phi}+ {\cal M}^{2}\phi = 0,\quad{\rm with}\quad {\cal M}^{2}\equiv \frac{\lambda M^{2}_{\rm P}}{3\xi^{2}}\,. \label{ST_2.1}
\end{eqnarray}
It was mentioned in \cite{GarciaBellido:2008ab} that the backreaction of $\chi$ particle into the dynamics of the inflaton field will only be relevant once its occupation numbers have grown sufficiently and can in principle suppress the resonant particle production. However, in this investigation, we will assume that the process of parametric resonance is at the stage in which the back reaction of the created particles can be neglected. Nevertheless, we will carefully examine the back reaction effect on the model by closely following the work studied in \cite{Kofman:1997yn} and will leave this for future investigation.
\begin{figure}[t]
\begin{center}		
\includegraphics[width=1\linewidth]{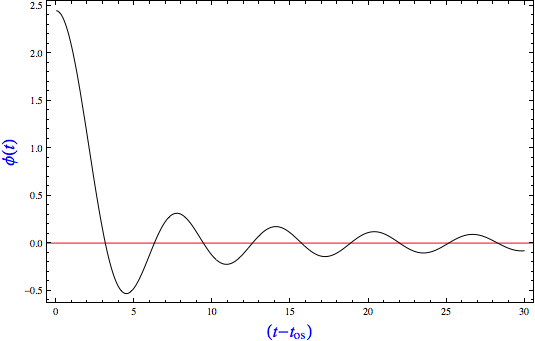}
\caption{We plot the approximate solution of the field $\phi(t)$ as given in Eq.(\ref{ST_2.21}). The value of the scalar field here is measured in units of $M_{\rm P}$ and time is measured in units of ${\cal M}^{-1}$. \label{phios}}
\end{center}
\end{figure}
In order to obtain the solution of the above equation, we use for a power-law evolution $a \propto t^{p}$ and the equation motion becomes
\begin{eqnarray}
t^{2}\ddot{\phi}+3pt\dot{\phi} + t^{2}{\cal M}^{2}\phi = 0\,. \label{ST_2.11}
\end{eqnarray}
The general solution of the effective equation of $\phi$ can be expressed in terms of the Bessel functions as
\begin{eqnarray}
\phi(t) = \frac{1}{({\cal M}t)^{\nu}}\Big[AJ_{+\nu}({\cal M}t) + BJ_{-\nu}({\cal M}t)\Big]\,, \label{ST_2.111}
\end{eqnarray}
with $A$ and $B$ constants depending on the initial conditions at the end of inflation, and $J_{\pm\nu}({\cal M}t)$ Bessel functions of order $\pm \nu$, with $\nu = (3p - 1)/2$. Note that for a reasonable power index, we use $p = 2/3$ for matter and $p = 1/2$ for radiation. Moreover, the second term on the right-hand side of Eq.(\ref{ST_2.111}) diverges in the limit ${\cal M}t \rightarrow 0$ and therefore should be neglected on physical grounds. The physical solution to this equation is then simply given by
\begin{eqnarray}
\phi(t) = A\,({\cal M}t)^{\frac{-(3p-1)}{2}}J_{\frac{(3p-1)}{2}}({\cal M}t)\,, \label{ST_2.1111}
\end{eqnarray}
where we have used the large argument expansion of fractional Bessel functions \cite{Abramowitz} such that ${\cal M}t\gg 1$. For the large argument expansion, the physical solution can be approximated by a cosinusoidal function \cite{GarciaBellido:2008ab}:
\begin{eqnarray}
\phi(t) \simeq A\,({\cal M}t)^{\frac{-(3p)}{2}}\cos\left({\cal M}(t-t_{\rm os}) - (3p\pi/4)\right)\,. \label{ST_2.12}
\end{eqnarray}
Here we can choose a constant $A$ by considering the oscillatory behaviour which starts just at the end of inflation, i.e. $\phi(t=t_{0}) = \phi_{\rm end} = \sqrt{6} M_{\rm P}\log(4/3)^{1/4}$. However, before further studying Eq.(\ref{ST_2.12}), it would be of great interest in examining the effective equation of state during the preheating phase. As already suggested in \cite{DeCross:2015uza} by using the virial theorem, the equation of state parameter $(\omega)$ for the system to background $(\varphi^{I})$ order is given by
\begin{eqnarray}
\omega \equiv \frac{\mathbb{p}}{\rho} = \frac{\dot{\psi}^{2} - 2{\cal U}}{\dot{\psi}^{2} + 2{\cal U}}\,, \label{omega}
\end{eqnarray}
which in turn $\mathbb{p}$ and $\rho$ the pressure and the energy density, and $\dot{\psi}^{2}\equiv {\cal G}_{IJ}{\dot\varphi}^{I}{\dot\varphi}^{J}$. Here we will estimate the parameter $\omega$ when the background fields begin to oscillate. As invetigated in \cite{DeCross:2015uza}, the actual dynamics of the system interpolates over the first few oscillations between matter-dominated and radiationed-domiated effective equation of state. 

More precisely, for large non-minimal coupling, the averaged parameter $\omega_{\rm avg}$ spends more times around $\omega_{\rm avg}\approx 0$ corresponding to $p=2/3$ for matter-dominated behavior as the universe continues to expand; while at late time the system behaves like radiation, that is to say $\omega_{\rm avg}\approx 1/3$ corresponding to $p=1/2$ for this case. In the present examination, we will study the dynamics of the oscillation by considering $p=2/3$ during the early time of the preheating phase. In this case, hence, we obtain the physical solution of Eq.(\ref{ST_2.12}) as
\begin{eqnarray}
\phi(t) = \frac{\phi_{\rm end}}{{\cal M}(t-t_{\rm os})}\sin({\cal M}(t-t_{\rm os}))\,, \label{ST_2.21}
\end{eqnarray}
where $t_{\rm os}$ is a time when the oscillating phase begins. In Fig.(\ref{phios}), we plot the evolution of $\phi(t)$ as governed by the approximate equation of motion of Eq.(\ref{ST_2.11}) with $p=2/3$, and we see that the amplitude of the first oscillation drops by a factor of ten during the first oscillation. From (\ref{EoMchi}), the equation of motion for the field $\chi$ becomes
\begin{eqnarray}
\ddot{\chi}+3H\dot{\chi} - \frac{1}{a^{2}}\nabla^{2}\chi +\Big[m^{2}_{\chi} + \sqrt{\frac{2}{3}}\frac{g^{2}M_{\rm P}\phi}{\xi}\Big] \chi = 0\,.  \label{ST_2.2}
\end{eqnarray}
Expanding the scalar fields $\chi$ in terms of the Heisenberg representation as
\begin{eqnarray}
\chi(t,{\bf x}) \sim \int \left(a_{k}\chi_{k}(t)e^{-i{\bf k}\cdot{\bf x}} + a^{\dagger}_{k}\chi^{*}_{k}(t)e^{i{\bf k}\cdot{\bf x}}\right)d^{3}{\bf k} \,, \label{mospace}
\end{eqnarray}
where $a_{k}$ and $a^{\dagger}_{k}$ are annihilation and creation operators, we find that $\chi_{k}$ obeys the following equation of motion:
\begin{eqnarray}
\ddot{\chi}_{k}+3H\dot{\chi}_{k} +\Big[ \frac{k^{2}}{a^{2}} + m^{2}_{\chi} + \sqrt{\frac{2}{3}}\frac{g^{2}M_{\rm P}\phi}{\xi}\Big]\chi_{k} = 0\,. \label{Pot211}
\end{eqnarray}
Fourier transforming this equation and rescaling the field by $Y_{k} = a^{3/2}\chi_{k}$ yields
\begin{eqnarray}
\ddot{Y}_{k} + \omega^{2}_{k}Y_{k} = 0\,, \label{Pot2111}
\end{eqnarray}
where a time dependent frequency of $Y_{k}$ is given by
\begin{eqnarray}
\omega^{2}_{k} = \frac{k^{2}}{a^{2}} + m^{2}_{\chi}+ \Big[\sqrt{\frac{2}{3}}\frac{g^{2}\phi_{\rm end}M_{\rm P}}{\xi}\frac{1}{T(t)}\sin(T(t))\Big]\,, \label{Pot20}
\end{eqnarray}
\begin{figure}[t]
\begin{center}		
\includegraphics[width=0.95\linewidth]{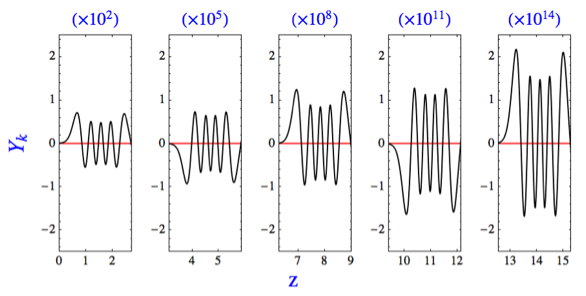}
\includegraphics[width=0.95\linewidth]{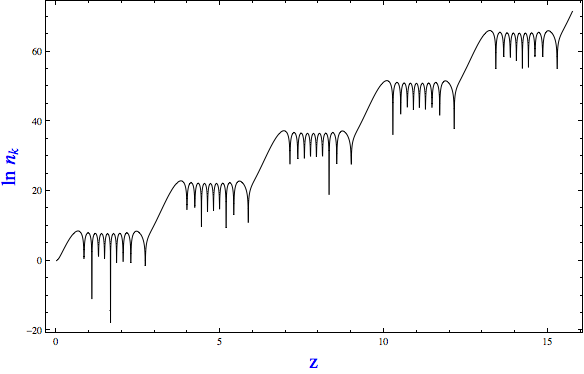}
\caption{By taking $k = 5{\cal M}\,(=m_{\chi}), q\sim 10^{2}$, the upper plot shows the amplification of the real part of the eigenmode $Y_{k}(z)$. The exponents show the order of magnitude for each given mode of fluctuations. We also see for each oscillation of the field $\phi(t)$ that the field $Y_{k}$ oscillates many times. The lower plot shows the logarithm of the comoving particle number density, $n_{k}(z)$, calculated with formula (\ref{growth}). The basic finding is that the number of particles grows exponentially, $\ln n_{k} \approx 2\mu_{k}z$. Here $z$ is measured in units of $2\pi/{\cal M}$. \label{Ynk}}
\end{center}
\end{figure}
with $T(t)={\cal M}(t-t_{\rm os})$. Certainly, Eq.(\ref{Pot2111}) describes an oscillator with a periodically changing frequency $\omega_{k}=k^{2}/a^{2} + m^{2}_{\chi}+ \Big[\sqrt{\frac{2}{3}}\frac{g^{2}\phi_{\rm end}M_{\rm P}}{\xi}\frac{1}{T(t)}\sin(T(t))\Big]$. With $a=1$, the physical momentum ${\bf p}$ coincides with ${\bf k}$ for Minkowski space such that $k=\sqrt{{\bf k}^{2}}$. The periodicity of Eq.(\ref{Pot2111}) may lead to the parametric resonance for modes with certain values of $k$. In order to examine this behavior, we will introduce a new variable, $z$, defined by ${\cal M}(t-t_{\rm os}) = 2z - \pi/2$. In the Minkowski space for which we neglect the expansion of the universe taking $a(t)=1$, Eq. (\ref{Pot2111}) turns to the standard Mathieu equation \cite{Mathieu}:
\begin{eqnarray}
\frac{d^{2}Y_{k}}{dz^{2}} + \left(A_{k} - 2q\cos(2z)\right)Y_{k} = 0\,, \label{Mathieu}
\end{eqnarray}
where
\begin{eqnarray}
A_{k} = \frac{4}{{\cal M}^{2}}\left(k^{2} + m^{2}_{\chi}\right)\,,\,\, q = \frac{6\sqrt{2}g^{2}}{\sqrt{3}\beta}\frac{\phi_{\rm end}}{T(t)} \,, \label{Matheiu2}
\end{eqnarray}
where $\beta=\lambda M_{\rm P}/\xi$. In general, the parameters $A_{k}$ and $q$ control the strength of parametric resonance. This can be described by a stability-instability chart of the Mathieu equation, see for example \cite{Kofman:1997yn}. As already mentioned above, the parameters $A_{k}$ and $q$ are constant in the Minkowski space. An important feature of solution of Eq.(\ref{Pot2111}) is the existence of an exponential instability $Y_{k} \propto \exp(\mu^{(n)}_{k}z)$ implying that $Y_{k}$ grows exponentially with a growth index $\mu_{k}$. In order to guarantee enough efficiency for the particle production, the Mathieu equation's parameters should satisfy the broad-resonance condition, i.e. $q\gg 1$. If this is the case, a broad resonance can possibly occur for a wide range of the parameter spaces and momentum modes. 

Therefore, in order to satisfy a broad resonance condition, we discover that $q\sim 12 g^{2}\xi/\lambda \gg 1$ implying $g^{2}/\lambda\gg 8.33\times 10^{-6}$. For a strongly coupled theory we expect $\lambda$ to be of the order of unity and in order to obtain successful inflation $\xi \sim {\cal O}(10^{4})$ is initially required. In this case, we can estimate the coupling $g\gg 3.0\times 10^{-3}$. The existence of an exponential instability $\chi_{k} \propto \exp(\mu^{(n)}_{k}z)$ causes an exponential growth of occupation numbers of quantum fluctuations $n_{k}(t) \propto \exp(2\mu^{(n)}_{k}z)$ that may be interpreted as particle production. 

Likewise, the growth of the modes $Y_{k}$ leads to the growth of the occupation numbers of the produced particles $n_{k}$. As suggested in \cite{Kofman:1997yn}, the number density of particles $n_{k}(t)$ with momentum ${\bf k}$ can be evaluated as the energy of that mode $\frac{1}{2}|\dot{Y}_{k}|^{2} + \frac{1}{2}\omega^{2}_{k}|Y_{k}|^{2}$ divided by the energy $\omega_{k}$ of each particle:
\begin{eqnarray}
n_{k}(z) = \frac{\omega_{k}}{2}\left(\frac{|\dot{Y}_{k}|^{2}}{\omega^{2}_{k}} + |Y_{k}|^{2}\right) - \frac{1}{2}\,. \label{growth}
\end{eqnarray}
Typically, in the simplest inflationary scenario including the one we are considering now, the value of the Hubble constant at the end of inflation is of the same order, but somewhat smaller, as the inflaton (effective) mass, ${\cal M}$. In order to clarify this, we can be even more concrete by evaluating the Hubble constant during the first oscillation. As shown in Fig.(\ref{phios}), during the first period of oscillation, the amplitudes of the field $\phi(t)$ drops to around $1/2$ of the reduced Planck mass, $M_{\rm P}$. We may also expect during this early phase of oscillation the field's kinetic energy to be roughly equal to its potential energy, and hence we can estimate the energy density of the field to be $\rho \sim {\cal M}^{2}\phi^{2}\sim \frac{1}{4}{\cal M}^{2}M^{2}_{\rm P}$. This approximation allows us to further estimate the Hubble constant and we find that the Hubble rate would then be $H=\sqrt{\frac{1}{3M^{2}_{\rm P}}\rho}\sim{\cal M}/(2/\sqrt{3}) \sim 0.3{\cal M}$. Notice that the estimate for $H/{\cal M} \sim 0.3$ is consistent with the exact treatment found for this exact same approach in Ref.\cite{DeCross:2015uza}, Fig.(8), in the limit $\xi_{\phi}\gg 1$.

It would be noticed that regarding the Mathieu equation there are instability bands in which the modes $Y_{k}$ grows exponentially with the growth index $\mu_{k} = q/2$. As we've mentioned above, these inability bands depends on the parameters $A_{k}$ and $q$. In order to warrant enough efficiency of the particle production, the Mathieu equation must satisfy the broad-resonance condition for which the conditions $A_{k}\simeq l^{2}$ where $l^{2} = 1,2,3,...,$ and $q\gg 1$ should be satisfied. Nevertheless, in general, the parameter $q$ decreases with time. Therefore it must take a large enough initial value.

Finally, we consider in this section the limiting case when the parameter $q$ takes a large value and make a plot the time evolution of fluctuations $Y_{k}$. We consider the typical resonance of particle production by taking $k = 5{\cal M}$ and $q\simeq 10^{2}$. The upper plot of Fig.(\ref{Ynk}) shows the amplification of the real part of the eigenmode $Y_{k}(z)$. Here the exponents show the order of the magnitude for each given mode of fluctuations. Apparently, the amplitude of the fluctuation for the second mode is much larger than that of the first one; while the third one is much larger than those of the first-two modes, and so forth. Moreover, the lower plot shows the logarithm of the comoving particle number density, $n_{k}(z)$, calculated with formula (\ref{growth}). The basic finding is that the number of particles grows exponentially, $\ln n_{k} \approx 2\mu_{k}z$. Here $z$ is measured in units of $2\pi/{\cal M}$. The jump of the number of particle density occur only near $z=z_{*}$ when the amplitude of the inflaton field crosses zero, i.e. $\phi(t=t_{*}) = 0$.

%%%%%%%%%%%%%%%%%%%%%%%
\section{V. Discussion and outlook}
%%%%%%%%%%%%%%%%%%%%%%%
We have investigated the production of particles due to parametric resonances in model of inflation in which the lightest composite state, which are expected to be the most important ones for collider phenomenology, stemming from the MWT theory. We studied a model of a composite inflaton field $\phi$ coupled to another scalar field $\chi$ with the interaction term $g^{2}\phi^{2}\chi^{2}$. Particularly, in the Minkowski space, the stage of parametric resonance can be simply described by the Mathieu equation. Furthermore, we discovered that broad resonances can be typically achieved and potentially efficient in our model. 

In order to satisfy a broad resonance. $q\gg 1$, we discover that $q\sim 12 g^{2}\xi/\lambda$ implying $g^{2}/\lambda\gg 8.33\times 10^{-6}$. For a strongly coupled theory we expect $\lambda$ to be of the order of unity and in order to obtain successful inflation $\xi \sim {\cal O}(10^{4})$ is initially required. In this case, we can estimate the coupling $g\gg 3.0\times 10^{-3}$. Quite surprisingly, the condition for a broad resonance in our model is very similar to that of the $f(R)$ gravity model \cite{DeFelice:2010aj} in the sense that the non-minimal coupling must take a large value. 

From Fig.(\ref{Ynk}), we discover that particle production in our model is potentially efficient causing the number of particles $n_{k}$ in this process remarkably increases. Using $q \simeq 10^{2}$, evidently for each oscillation of the field $\phi(t)$, the basic finding is that the number of particles grows exponentially, $\ln n_{k} \approx 2\mu{k}z$. Note that we here neglected the expansion of the universe by taking $a =1$ in Eq. (\ref{Mathieu}) and assumed that $k = 5{\cal M}=m_{\chi}$.

Once again, in this exploratory study our approach is minimalistic rather than aiming at great generality. Since during the parametric resonance stage, the produced particles are far away from equilibrium. As a result, the study of the thermalization at the end of the parametric resonance regime is also of interest, and yet the backreaction effect of other particles into the dynamics of the inflaton is also interesting. However, we will leave these interesting topics for our future project.

\vskip 2pt
\paragraph*{\bf Acknowledgements.}
 
We thank the reviewers for very helpful suggestions. P.C. is financially supported by the Institute for the Promotion of Teaching Science and Technology (IPST) under the project of the \lq\lq Research Fund for DPST Graduate with First Placement\rq\rq\,under Grant No. 033/2557 and by the Thailand Research Fund (TRF) under the project of the \lq\lq TRF Grant for New Researcher\rq\rq\,under Grant No. TRG5780143. P.K. is financially supported by the National Science and Technology Development Agency (NSTDA) and the TRF under the Junior Science Talent Project (JSTP) with Grant No. JSTP-06-57-01E.

%\end{acknowledgements}
%%%%%%%%%%%%%%%%%%%%%%%%%%%%%%%%%%%%%%%%%%%%%%%%%%%%%%%%%%%%%%%%%%%%%%%%%%%%%%%

%%%%%%%%%%%%%%%%%%%%%%%%%%%%%%%%%%%%%%%%%%%%%%%%%%%%%%%%%%%%%%%%%%%%%%%%%%%%%%%


\begin{thebibliography}{99}

%%%%%%%%%%%Reheating%%%%%%%%%%%%
%\cite{Starobinsky:1980te}
\bibitem{Starobinsky:1980te} 
  A.~A.~Starobinsky, ``A New Type of Isotropic Cosmological Models Without Singularity,''
  Phys.\ Lett.\ B {\bf 91}, 99 (1980).

%\cite{Mukhanov:1981xt}
\bibitem{Mukhanov:1981xt} 
  V.~F.~Mukhanov and G.~V.~Chibisov, ``Quantum Fluctuation and Nonsingular Universe. (In Russian),''
  JETP Lett.\  {\bf 33}, 532 (1981) [Pisma Zh.\ Eksp.\ Teor.\ Fiz.\  {\bf 33}, 549 (1981)].

%\cite{Guth:1980zm}
\bibitem{Guth:1980zm} 
  A.~H.~Guth, ``The Inflationary Universe: A Possible Solution to the Horizon and Flatness Problems,''
  Phys.\ Rev.\ D {\bf 23}, 347 (1981).

%\cite{Linde:1981mu}
\bibitem{Linde:1981mu} 
  A.~D.~Linde, ``A New Inflationary Universe Scenario: A Possible Solution of the Horizon, Flatness, Homogeneity, Isotropy and Primordial Monopole Problems,''
  Phys.\ Lett.\ B {\bf 108}, 389 (1982).

%\cite{Albrecht:1982wi}
\bibitem{Albrecht:1982wi} 
  A.~Albrecht and P.~J.~Steinhardt, ``Cosmology for Grand Unified Theories with Radiatively Induced Symmetry Breaking,''
  Phys.\ Rev.\ Lett.\  {\bf 48}, 1220 (1982).
  
  %%%%%%%%%%%Observations%%%%%%%%%%%
  
  %\cite{Peiris:2003ff}
\bibitem{Peiris:2003ff} 
  H.~V.~Peiris {\it et al.} [WMAP Collaboration], ``First year Wilkinson Microwave Anisotropy Probe (WMAP) observations: Implications for inflation,''
  Astrophys.\ J.\ Suppl.\  {\bf 148}, 213 (2003)
  
  %\cite{Ade:2015lrj}
\bibitem{Ade:2015lrj} 
  P.~A.~R.~Ade {\it et al.} [Planck Collaboration], ``Planck 2015 results. XX. Constraints on inflation,''
  arXiv:1502.02114 [astro-ph.CO].
  
%%%%%%%%%%%%%%%%%%%%%%%%%%%%%%
  
%\cite{Bezrukov:2007ep}
\bibitem{Bezrukov:2007ep} 
  F.~L.~Bezrukov and M.~Shaposhnikov, ``The Standard Model Higgs boson as the inflaton,''
  Phys.\ Lett.\ B {\bf 659}, 703 (2008)

%\cite{Maleknejad:2011sq}
\bibitem{Maleknejad:2011sq} 
  A.~Maleknejad and M.~M.~Sheikh-Jabbari, ``Non-Abelian Gauge Field Inflation,''
  Phys.\ Rev.\ D {\bf 84}, 043515 (2011)
  
  %\cite{Germani:2009iq}
\bibitem{Germani:2009iq} 
  C.~Germani and A.~Kehagias, ``P-nflation: generating cosmic Inflation with p-forms,''
  JCAP {\bf 0903}, 028 (2009)
  
  %\cite{Koivisto:2009sd}
\bibitem{Koivisto:2009sd} 
  T.~S.~Koivisto, D.~F.~Mota and C.~Pitrou, ``Inflation from N-Forms and its stability,''
  JHEP {\bf 0909}, 092 (2009)
  
%\cite{Kaiser:2013sna}
\bibitem{Kaiser:2013sna} 
  D.~I.~Kaiser and E.~I.~Sfakianakis, ``Multifield Inflation after Planck: The Case for Nonminimal Couplings,''
  Phys.\ Rev.\ Lett.\  {\bf 112}, no. 1, 011302 (2014)
  
%\cite{Channuie:2011rq}
\bibitem{Channuie:2011rq}
  P.~Channuie, J.~J.~Joergensen and F.~Sannino, ``Minimal Composite Inflation,''
  JCAP {\bf 1105} (2011) 007

%\cite{Bezrukov:2011mv}
\bibitem{Bezrukov:2011mv} 
  F.~Bezrukov, P.~Channuie, J.~J.~Joergensen and F.~Sannino, ``Composite Inflation Setup and Glueball Inflation,''
  Phys.\ Rev.\ D {\bf 86}, 063513 (2012)

%\cite{Channuie:2012bv}
\bibitem{Channuie:2012bv}
  P.~Channuie, J.~J.~Jorgensen and F.~Sannino, ``Composite Inflation from Super Yang-Mills, Orientifold and One-Flavor QCD,''
  Phys.\ Rev.\ D {\bf 86} (2012) 125035
  
%\cite{Linde:2005ht}
\bibitem{Linde:2005ht} 
  A.~D.~Linde, ``Particle physics and inflationary cosmology,''
  Contemp.\ Concepts Phys.\  {\bf 5}, 1 (1990)
  
  \bibitem{Albrecht:1982}
  A. Albrecht, P. J. Steinhardt, M. S. Turner and F Wilczek, "Reheating an inflationary universe", Phys. Rev. Lett. 48 (1982) 1437
  
%\cite{Abbott:1982hn}
\bibitem{Abbott:1982hn} 
  L.~F.~Abbott, E.~Farhi and M.~B.~Wise, ``Particle Production in the New Inflationary Cosmology,''
  Phys.\ Lett.\ B {\bf 117}, 29 (1982).
  
  \bibitem{Starobinsky:1981} 
A. A. Starobinsky, in: Quantum Gravity, Proc. of the second seminar "quantum theory of gravity", (Moscow, 13-15 October 1981
  
%\cite{Perrier:2012nr}
\bibitem{Perrier:2012nr} 
  H.~Perrier, R.~Durrer and M.~Rinaldi, ``Explosive particle production in non-commutative inflation,''
  JHEP {\bf 1301}, 067 (2013)
  
   %%%%%%%%a resonance%%%%%%%
 %\cite{Traschen:1990sw}
\bibitem{Traschen:1990sw} 
  J.~H.~Traschen and R.~H.~Brandenberger, ``Particle Production During Out-of-equilibrium Phase Transitions,''
  Phys.\ Rev.\ D {\bf 42}, 2491 (1990).
  
 %\cite{Kofman:1997yn}
\bibitem{Kofman:1997yn} 
  L.~Kofman, A.~D.~Linde and A.~A.~Starobinsky, ``Towards the theory of reheating after inflation,''
  Phys.\ Rev.\ D {\bf 56}, 3258 (1997)
  
  %\cite{Shtanov:1994ce}
\bibitem{Shtanov:1994ce} 
  Y.~Shtanov, J.~H.~Traschen and R.~H.~Brandenberger, ``Universe reheating after inflation,''
  Phys.\ Rev.\ D {\bf 51}, 5438 (1995)
  
  %\cite{Kofman:1994rk}
\bibitem{Kofman:1994rk} 
  L.~Kofman, A.~D.~Linde and A.~A.~Starobinsky, ``Reheating after inflation,''
  Phys.\ Rev.\ Lett.\  {\bf 73}, 3195 (1994)
 
 %%%%%%%%Preheat%%%%%%%%%
 
 %\cite{Greene:1997fu}
\bibitem{Greene:1997fu} 
  P.~B.~Greene, L.~Kofman, A.~D.~Linde and A.~A.~Starobinsky, ``Structure of resonance in preheating after inflation,''
  Phys.\ Rev.\ D {\bf 56}, 6175 (1997)
 
 %\cite{Kaiser:1995fb}
\bibitem{Kaiser:1995fb} 
  D.~I.~Kaiser, ``Post inflation reheating in an expanding universe,''
  Phys.\ Rev.\ D {\bf 53}, 1776 (1996)
 
 %\cite{Son:1996uv}
\bibitem{Son:1996uv} 
  D.~T.~Son, ``Reheating and thermalization in a simple scalar model,''
  Phys.\ Rev.\ D {\bf 54}, 3745 (1996)
  
%\cite{Tsujikawa:1999jh}
\bibitem{Tsujikawa:1999jh} 
  S.~Tsujikawa, K.~i.~Maeda and T.~Torii, ``Resonant particle production with nonminimally coupled scalar fields in preheating after inflation,''
  Phys.\ Rev.\ D {\bf 60}, 063515 (1999)
  
%\cite{Tsujikawa:1999iv}
\bibitem{Tsujikawa:1999iv} 
  S.~Tsujikawa, K.~i.~Maeda and T.~Torii, ``Preheating with nonminimally coupled scalar fields in higher curvature inflation models,''
  Phys.\ Rev.\ D {\bf 60}, 123505 (1999)
  
  %\cite{GarciaBellido:2008ab}
\bibitem{GarciaBellido:2008ab} 
  J.~Garcia-Bellido, D.~G.~Figueroa and J.~Rubio, ``Preheating in the Standard Model with the Higgs-Inflaton coupled to gravity,''
  Phys.\ Rev.\ D {\bf 79}, 063531 (2009)
  
  %\cite{Copeland:1994vg}
\bibitem{Copeland:1994vg} 
  E.~J.~Copeland, A.~R.~Liddle, D.~H.~Lyth, E.~D.~Stewart and D.~Wands, ``False vacuum inflation with Einstein gravity,''
  Phys.\ Rev.\ D {\bf 49}, 6410 (1994)
  
  %\cite{Channuie:2013lla}
\bibitem{Channuie:2013lla} 
  K.~Karwan and P.~Channuie, ``Composite Inflation confronts BICEP2 and PLANCK,''
  JCAP {\bf 1406}, 045 (2014)
  
  %\cite{Channuie:2015ewa}
\bibitem{Channuie:2015ewa} 
  P.~Channuie, ``Composite Inflation in the light of 2015 Planck data,''
  arXiv:1510.05262 [hep-ph].
  
%\cite{Array:2015xqh}
\bibitem{Array:2015xqh} 
  P.~A.~R.~Ade {\it et al.} [BICEP2 and Keck Array Collaborations], ``Improved Constraints on Cosmology and Foregrounds from BICEP2 and Keck Array Cosmic Microwave Background Data with Inclusion of 95 GHz Band,''
  Phys.\ Rev.\ Lett.\  {\bf 116}, 031302 (2016)
   
%%%%%%%%EndReheat%%%%%%%%%%%%%%%

%\cite{Futamase:1987ua}
\bibitem{Futamase:1987ua} 
  T.~Futamase and K.~i.~Maeda,
  ``Chaotic Inflationary Scenario in Models Having Nonminimal Coupling With Curvature,''
  Phys.\ Rev.\ D {\bf 39}, 399 (1989).
  
%\cite{Mahajan:2013kra}
\bibitem{Mahajan:2013kra} 
  N.~Mahajan, ``Some remarks on nonminimal coupling of the inflaton,''
  Int.\ J.\ Mod.\ Phys.\ D {\bf 23}, no. 09, 1450076 (2014)
  
  %\cite{Steinwachs:2013tr}
\bibitem{Steinwachs:2013tr} 
  C.~F.~Steinwachs and A.~Y.~Kamenshchik,
  ``Non-minimal Higgs Inflation and Frame Dependence in Cosmology,''
  AIP Conf.\ Proc.\  {\bf 1514}, 161 (2012)

%\cite{Kaiser:2010ps}
\bibitem{Kaiser:2010ps} 
  D.~I.~Kaiser, ``Conformal Transformations with Multiple Scalar Fields,''
  Phys.\ Rev.\ D {\bf 81}, 084044 (2010)
  
  %\cite{Bassett:1997az}
\bibitem{Bassett:1997az} 
  B.~A.~Bassett and S.~Liberati,
  %``Geometric reheating after inflation,''
  Phys.\ Rev.\ D {\bf 58}, 021302 (1998)
  Erratum: [Phys.\ Rev.\ D {\bf 60}, 049902 (1999)]
  doi:10.1103/PhysRevD.60.049902, 10.1103/PhysRevD.58.021302
  [hep-ph/9709417].
  %%CITATION = doi:10.1103/PhysRevD.60.049902, 10.1103/PhysRevD.58.021302;%%
  %54 citations counted in INSPIRE as of 13 Jul 2016
  
  %\cite{Tsujikawa:2002nf}
\bibitem{Tsujikawa:2002nf} 
  S.~Tsujikawa and B.~A.~Bassett,
  %``When can preheating affect the CMB?,''
  Phys.\ Lett.\ B {\bf 536}, 9 (2002)
  doi:10.1016/S0370-2693(02)01813-0
  [astro-ph/0204031].
  %%CITATION = doi:10.1016/S0370-2693(02)01813-0;%%
  %42 citations counted in INSPIRE as of 13 Jul 2016
  
  %\cite{Amin:2014eta}
\bibitem{Amin:2014eta} 
  M.~A.~Amin, M.~P.~Hertzberg, D.~I.~Kaiser and J.~Karouby,
  %``Nonperturbative Dynamics Of Reheating After Inflation: A Review,''
  Int.\ J.\ Mod.\ Phys.\ D {\bf 24}, 1530003 (2014)
  doi:10.1142/S0218271815300037
  [arXiv:1410.3808 [hep-ph]].
  %%CITATION = doi:10.1142/S0218271815300037;%%
  %53 citations counted in INSPIRE as of 13 Jul 201
  
  %\cite{DeCross:2015uza}
\bibitem{DeCross:2015uza} 
  M.~P.~DeCross, D.~I.~Kaiser, A.~Prabhu, C.~Prescod-Weinstein and E.~I.~Sfakianakis,
  %``Preheating after Multifield Inflation with Nonminimal Couplings, I: Covariant Formalism and Attractor Behavior,''
  arXiv:1510.08553 [astro-ph.CO].
  %%CITATION = ARXIV:1510.08553;%%
  %6 citations counted in INSPIRE as of 13 Jul 2016
  
  %\cite{Sannino:2004qp}
\bibitem{Sannino:2004qp} 
  F.~Sannino and K.~Tuominen, ``Orientifold theory dynamics and symmetry breaking,''
  Phys.\ Rev.\ D {\bf 71}, 051901 (2005)
  
%\cite{Hong:2004td}
\bibitem{Hong:2004td} 
  D.~K.~Hong, S.~D.~H.~Hsu and F.~Sannino, ``Composite Higgs from higher representations,''
  Phys.\ Lett.\ B {\bf 597}, 89 (2004)
  
 %\cite{Dietrich:2005wk}
\bibitem{Dietrich:2005wk} 
  D.~D.~Dietrich, F.~Sannino and K.~Tuominen,  ``Light composite Higgs and precision electroweak measurements on the Z resonance: An Update,''
  Phys.\ Rev.\ D {\bf 73}, 037701 (2006)
  
  %\cite{Dietrich:2005jn}
\bibitem{Dietrich:2005jn} 
  D.~D.~Dietrich, F.~Sannino and K.~Tuominen, ``Light composite Higgs from higher representations versus electroweak precision measurements: Predictions for CERN LHC,''
  Phys.\ Rev.\ D {\bf 72}, 055001 (2005)

  %\cite{Hill:1991jc}
\bibitem{Hill:1991jc} 
  C.~T.~Hill and D.~S.~Salopek, ``Calculable nonminimal coupling of composite scalar bosons to gravity,''
  Annals Phys.\  {\bf 213}, 21 (1992).
  
 %%%%%%baryogenesis%%%%%%  
  %\cite{Kolb:1996jt}
\bibitem{Kolb:1996jt} 
  E.~W.~Kolb, A.~D.~Linde and A.~Riotto, ``GUT baryogenesis after preheating,''
  Phys.\ Rev.\ Lett.\  {\bf 77}, 4290 (1996)
 
\bibitem{Mathieu}
N. McLachlan, ``Theory and Applications of Mathieu 
Functions,'' (Oxford Univ. Press, Clarendon, 1947)

\bibitem{Abramowitz}
M. Abramowitz and I. A. Stegun, ``Handbook of Mathematical Functions with Formulas, Graphs, and Mathematical
Tables,'' New York, Dover (1972)
  
  %\cite{DeFelice:2012wy}
\bibitem{DeFelice:2012wy} 
  A.~De Felice, K.~Karwan and P.~Wongjun, ``Reheating in 3-form inflation,''
  Phys.\ Rev.\ D {\bf 86}, 103526 (2012)
  
  %\cite{DeFelice:2010aj}
\bibitem{DeFelice:2010aj} 
  A.~De Felice and S.~Tsujikawa, ``f(R) theories,''
  Living Rev.\ Rel.\  {\bf 13}, 3 (2010)
 

\end{thebibliography}
\end{document}